\documentclass[sigconf]{acmart}
\usepackage[ruled,linesnumbered]{algorithm2e}
\usepackage{amsfonts}
\usepackage{amsmath}
\usepackage{amsthm}
\usepackage{bm}
\usepackage{soul}
\usepackage[utf8]{inputenc}
\usepackage[T1]{fontenc}
\usepackage{lmodern}
\pdfmapfile{=tfrupee.map}

\AtBeginDocument{%
  \providecommand\BibTeX{{%
    \normalfont B\kern-0.5em{\scshape i\kern-0.25em b}\kern-0.8em\TeX}}}
\begin{document}

%%
%% The "title" command has an optional parameter,
%% allowing the author to define a "short title" to be used in page headers.
\title{Proof of Learning (PoLe): Empowering Machine Learning with Consensus Building on Blockchains}

%%
%% The "author" command and its associated commands are used to define
%% the authors and their affiliations.
%% Of note is the shared affiliation of the first two authors, and the
%% "authornote" and "authornotemark" commands
%% used to denote shared contribution to the research.
\author{Yixiao Lan}
\affiliation{%
  \institution{Northeastern University, China}
}
\email{yixiaolan@foxmail.com}
%\orcid{1234-5678-9012}
\author{Yuan Liu}
%\authornotemark[1]

\affiliation{%
  \institution{Northeastern University, China}
}
\email{liuyuan@swc.neu.edu.cn}
\author{Boyang Li}
\affiliation{%
  \institution{Alibaba-NTU Singapore Joint Research Institute}
 }
\email{lily.liboyang@ntu.edu.sg}

%\author{Valerie B\'eranger}
%\affiliation{%
%  \institution{Inria Paris-Rocquencourt}
%  \city{Rocquencourt}
%  \country{France}
%}
%
%\author{Aparna Patel}
%\affiliation{%
% \institution{Rajiv Gandhi University}
% \streetaddress{Rono-Hills}
% \city{Doimukh}
% \state{Arunachal Pradesh}
% \country{India}}
%
%\author{Huifen Chan}
%\affiliation{%
%  \institution{Tsinghua University}
%  \streetaddress{30 Shuangqing Rd}
%  \city{Haidian Qu}
%  \state{Beijing Shi}
%  \country{China}}
%
%\author{Charles Palmer}
%\affiliation{%
%  \institution{Palmer Research Laboratories}
%  \streetaddress{8600 Datapoint Drive}
%  \city{San Antonio}
%  \state{Texas}
%  \postcode{78229}}
%\email{cpalmer@prl.com}
%
%\author{John Smith}
%\affiliation{\institution{The Th{\o}rv{\"a}ld Group}}
%\email{jsmith@affiliation.org}
%
%\author{Julius P. Kumquat}
%\affiliation{\institution{The Kumquat Consortium}}
%\email{jpkumquat@consortium.net}

%%
%% By default, the full list of authors will be used in the page
%% headers. Often, this list is too long, and will overlap
%% other information printed in the page headers. This command allows
%% the author to define a more concise list
%% of authors' names for this purpose.
%\renewcommand{\shortauthors}{Anonomous, et al.}

%%
%% The abstract is a short summary of the work to be presented in the
%% article.
\begin{abstract}
  The progress of deep learning (DL), especially the recent development of automatic design of networks, has brought unprecedented performance gains at heavy computational cost. On the other hand, blockchain systems routinely perform a huge amount of computation that does not achieve practical purposes in order to build Proof-of-Work (PoW) consensus from decentralized participants. In this paper, we propose a new consensus mechanism, Proof of Learning (PoLe), which directs the computation spent for consensus toward optimization of neural networks (NN). In our mechanism, the training/testing data are released to the entire blockchain network (BCN) and the consensus nodes train NN models on the data, which serves as the proof of learning. When the consensus on the BCN considers a NN model to be valid, a new block is appended to the blockchain.  We experimentally compare the PoLe protocol with Proof of Work (PoW) and show that PoLe can achieve a more stable block generation rate, which leads to more efficient
  transaction processing. We also introduce a novel cheating prevention mechanism, Secure Mapping Layer (SML), which can be straightforwardly implemented as a linear NN layer. Empirical evaluation shows that 
  SML can detect cheating nodes at small cost to the predictive performance.
\end{abstract}

%%
%% The code below is generated by the tool at http://dl.acm.org/ccs.cfm.
%% Please copy and paste the code instead of the example below.
%%



%%
%% Keywords. The author(s) should pick words that accurately describe
%% the work being presented. Separate the keywords with commas.

%% A "teaser" image appears between the author and affiliation
%% information and the body of the document, and typically spans the
%% page.
%\begin{teaserfigure}
%  \includegraphics[width=\textwidth]{sampleteaser}
%  \caption{Seattle Mariners at Spring Training, 2010.}
%  \Description{Enjoying the baseball game from the third-base
%  seats. Ichiro Suzuki preparing to bat.}
%  \label{fig:teaser}
%\end{teaserfigure}

%%
%% This command processes the author and affiliation and title
%% information and builds the first part of the formatted document.
\maketitle

\section{Introduction}
Recently, significant performance gain in deep learning has been derived from scaling up the network and training data  \cite{mahajan2018exploring,Radford2019:GPT-2,Devlin2019:BERT,yalniz2019billionscale} and automatic design of neural network (NN) architectures \cite{So2019:EvolvedTransformer,Real2019:AmoebaNet,Xin2019}. These trends have created an ever-growing demand for computational power. For example, the largest GPT-2 model \cite{Radford2019:GPT-2} has a total of 1,542 million parameters. The evolution of the transformer model from \cite{So2019:EvolvedTransformer} required 32,623 TPU hours or 2,192,960 P100 GPU hours. Such compute requirements render model optimization unrealistic for academic researchers and even small-sized enterprises and hinder wide adoption of artificial intelligence technologies. The energy consumption leaves large carbon footprint and non-negligible environmental impact \cite{strubell2019energy}. 

A blockchain network (BCN) is a decentralized distributed system where participants do not trust each other but are able to collectively maintain a consistent database or ledger. Pioneered by the Bitcoin system~\cite{nakamoto2008bitcoin}, BCN is now used in novel financial services such as crytocurrency and smart contracts. It has also been envisioned to have major roles in applications that require permanent and tampering-resistant data storage. 

However, a major drawback of BCN is that, due to the Proof-of-Work (PoW) mechanism that builds consensus from distrustful nodes, it consumes massive amount of compute and energy. According to the Cambridge Bitcoin Electricity Consumption Index\footnote{https://www.cbeci.org/}, at the time of writing, the annualized electricity consumption of Bitcoin mining exceeds that of Austria. Most of the energy is spent on calculating the nonce of hash functions, which serves no real purposes other than being difficult to compute. In this paper, we direct the computation and energy spent on blockchain consensus to the practical function of training machine learning models. 

A consensus mechanism on the BCN specifies how the participants determine an increment to the data ledger, or a block, is valid. Only vaid blocks are added to the ledger. Therefore, the consensus mechanism is key to the consistency and integrity of the ledger. In the popular Proof-of-Work (PoW) mechanism, participants compete to solve an asymmetric hash puzzle that is difficult to solve but easy to verify. One solution is a nonce $x$ whose hash $h(x)$ begins with a specific number of zeros \cite{conti2018survey}. Once the nonce is found, other participants can easily verify if the solution is correct. The problem solver is then allowed to append a new block to the blockchain. Later blocks on the chain contain the hash of the previous block. Tampering with an old block hence requires regenerating all subsequent blocks, which is prohibitively expensive.

Similar to PoW, the proposed Proof-of-Learning (PoLe) consensus mechanism contains two main constituents: an assymetric puzzle and a tampering prevention technique that links adjacent blocks. For the  assymmetric puzzle, we rely on the non-convex optimization of large neural networks, based on the realization that it is 
NP-hard to find a set of network weights that achieve a given training accuracy, but easy to verify if a given set of weights can achieve a given training or test accuracy. For tampering prevention, we propose a functional encryption scheme, whose decryption can be implemented as a linear layer in the neural network called the Secure Mapping Layer (SML). The SML projects input feature vectors onto specific directions that depend on the hash of the previous block. This technique prevents participants from reusing exsting solutions in order to gain unfair advantages. Furthermore, it allows the PoLe BCN, in the absense of useful machine learning work, to gracefully degrade to PoW by repeatedly commission machine learning tasks for the sole purpose of new block generation. 

The proposed PoLe-based BCN provides a platform on which users may commission a neural network model that meet their requirements. The BCN contains two types of participants, data nodes, which announce tasks with a reward, and consensus nodes or miners, which work to solve announced tasks. 
After a data node announces a task, consensus nodes may accept it and seek a model that meets the announced minimum training accuracy. After receiving a valid solution, which meets the training accuracy criterion, the data node releases the test set. The consensus nodes then collectively select the solution with the highest generalization performance and distribute the reward accordingly. 
Thus, a PoLe-based blockchain can serve as a decentralized database and a machine-learning platform simultaneously. 

The contributions of this work are summarized as follows.
\begin{itemize}
    \item We propose a new consensus algorithm on blockchain called Proof of Learning (PoLe), which channels the otherwise wasted computational power to the practical purpose of training neural network models. 
    \item In order to maintain the integrity of the blockchain ledger, we propose a data encryption scheme that uses a linear NN layer to link adjacent blocks together. Empirical evidence shows that the encryption detects tampering behavior without significantly degrading predictive performance.
    \item We provide a mechanism design that encourages two types of participants, data nodes and consensus nodes, to collaborate, while the consensus nodes compete among themselves. The mechanism rewards the consensus nodes for the best generalization performance.  
\end{itemize}

%In this paper, we propose a new consensus algorithm called PoLe and its corresponding blockchain model. This consensus algorithm transforms the workload of finding nouce and calculating hash value in POW algorithm into the workload of training deep-learning model, and uses the trained model as the proof of workload to ensure the data security of blockchain system. At the same time, when users upload data to blockchain network in order to train their own model, it means that users' privacy and identity will be exposed. In order to ensure the privacy of users' data, we propose a method to train deep learning model on encrypted data. In section  4 and 5, we describe the PoLe algorithm and its corresponding blockchain model in detail. In the section 7, we use experiments to verify the feasibility of the algorithm and test the performance of the deep learning model trained on encrypted data by the proposed method.

% Machine learning, especially deep-learning, is also widely used in many fields, such as business and manufacturing. Deep learning uses labeled data to train deep learning model in the way of iterative updating of weights. A well structured deep learning model can fit any function, which makes deep-learning show strong productivity in solving specific problems. However, the training of deep learning model usually consumes a lot of computing resources. With the extensive application of deep learning, the demand of computing power will be more and more. 

\section{Ralated work}
\subsection{Blockchain and Consensus Mechanisms}
The technology of blockchain is a distributed paradigm to boost  the internet of value~\cite{Tapscott2019Blockchain}, which has gained tremendous momentum in the past decade. Blockchain enables distributed parties who do not fully trust each other to maintain a shared and consistent ledger. Blockchain networks (BCNs) are distinguished by unique characteristics including decentralization, traceability, transparency, tampering-resistance, and programmability~\cite{DataProcessingViewBlk2018}.

Consensus algorithm is a method to achieve data consistency among multiple distributed nodes. The most classical and commonly used consensus algorithm is proof of work (PoW) proposed in Bitcoin white paper \cite{nakamoto2008bitcoin}, which was ever proposed in filtering spam.
In the Bitcoin network, the PoW algorithm lets each node repeatedly generate nonces when ``mining'' a valid block, until the hash value calculated by the nonce and content of the current and previous blocks less than a target value is worked out, which will be used as the hash value of the current block. The blockchain model always considers the longest blockchain to be effective. When a node wants to modify the content of a previous block and make this content effective, it needs to recalculate the hash value of all blocks after that block and make the length of the modified blockchain longer than the unmodified chain. The computational power required by this process will be prohibitively high, which guarantees the data security and data consistency of the blockchain systems.

Motivated to reduce the compute demand of PoW, proof of stake (PoS)~\cite{king2012ppcoin} and distributed proof of stake (DPoS)~\cite{larimer2014delegated} have been proposed as alternatives to PoW. PoS dynamically adjusts the PoW calculation difficulty of each node according to the amount and time of tokens held by different nodes, so as to form a situation where the nodes with more token ages are more likely to take the lead in calculating available blocks. Although PoS reduces the need for compute, it places trust in participants with a large amount of tokens, which may collude to become centers and arbitrators of the network. Centralization may erode trust in the fairness of the consensus. This concern may have played a role in the slow transition of the Ethereum platform, which is attempting to transit from PoW to PoS.

Other alternative consensus mechanisms such as credit-based PoW~\cite{huang2019towards}, proof of reputation (PoR) \cite{PoRyuan2019} and proof of negotiation (PoN) \cite{DBLP:journals/fgcs/FengZCZZ20}, have been also proposed in the literature.  PoR studies a two-chain architecture to construct the reputation of nodes in a separate chain and the next block generator is determined by the reputation chain~\cite{PoRyuan2019}. In PoN, the trustworthiness of miners are evaluated and a random-honest miner is selected based on negotiation rules. The PoN investigates parallel multi-block creation method to achieve high efficiency than traditional consensus mechanisms in one-by-one block creation~\cite{DBLP:journals/fgcs/FengZCZZ20}.   To date, PoW remains the most popular and widely accepted choice~\cite{gervais2016security}.

%  PoW is one of the most widely used consensus algorithms. It is used in bitcoin, the first blockchain model in the world. It has many advantages. It uses simple and repeated calculation methods to ensure the honest work of each node, and achieves the whole network consensus through workload, so as to ensure data security. 
 
% PoW consumes a lot of computing power to perform repeated hash calculation, and these calculations are only used to prove that they have done the calculation, and the calculation itself is meaningless. For PoS, although it is an upgraded version of PoW, it still needs to consume computing power to make meaningless hash calculation. Moreover, when some nodes have more and more tokens, the whole system will incline to centralization. 
%  On the other hand, privacy technology is one of the important technologies to protect blockchain. Many new encryption algorithms are used in the blockchain to ensure the privacy of user data and user information, such as zero knowledge proof and ring signature algorithm.  These algorithms have different characteristics and guarantee the privacy of different parts of the blockchain system.

Replacing hash calculation, the analog Hamiltonian optimisers based PoW is proposed in \cite{kalinin2018blockchain} such that the blockchain system can generate blocks in seconds to save energy. Some afforts have been made to direct the computational power in Bitcoin to scientific computing, such as Primecoin\footnote{https://primecoin.io/about.php}. In PrimeCoin network, a prime PoW is designed to generates a special form of prime number chains of interest to mathematical research. In this work, we aims to apply the computational resources in PoW in a general case of deep learning model training.
%\cite{kalinin2018blockchain} hl {uses blockchain to solve other optimization problems.} 
\begin{figure*}[!ht]
\centering
	\includegraphics[width=0.9\textwidth]{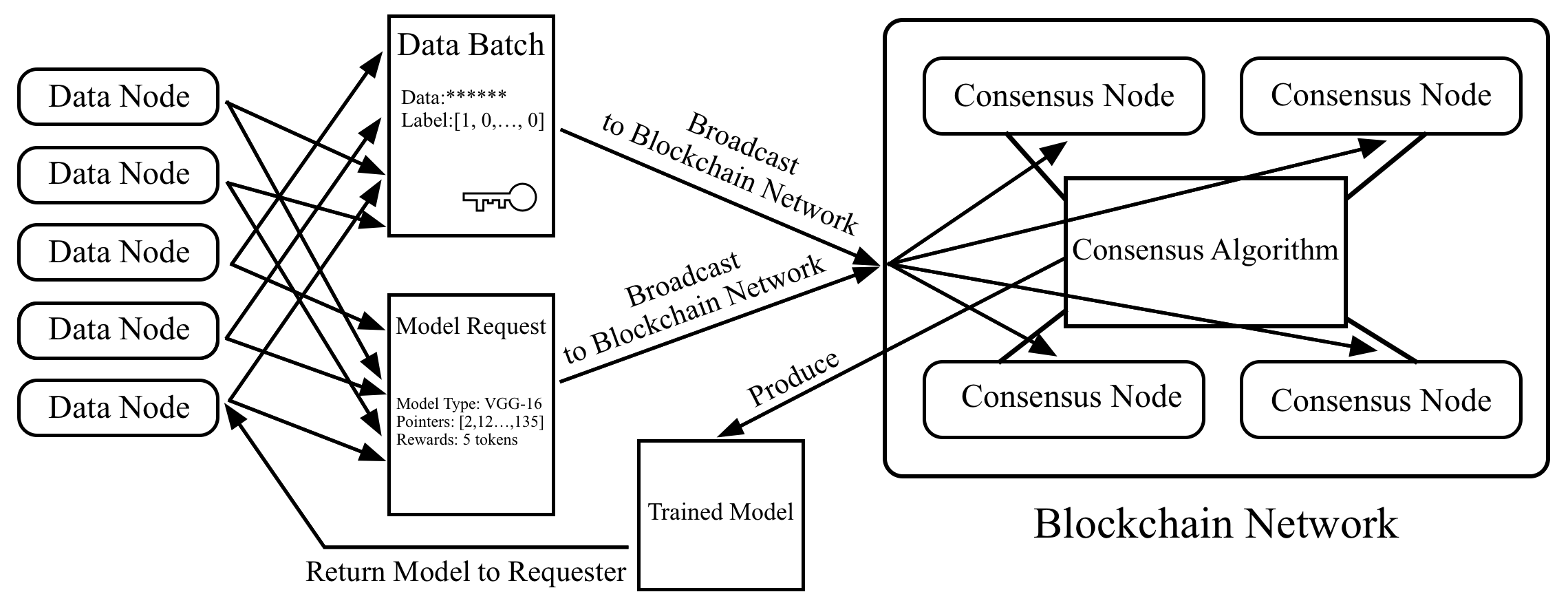}
	\caption{The Overview of the Proposed System}
	\label{FigSystem}
\end{figure*}
\subsection{Federated Learning}
%Federated learning  aims to achieve secure machine learning on distributed systems. 
The proposed blockchain system bears some resemblance to federated learning \cite{yang2019federated,konen2016federated}, where various participants with different data collaborate in a secure and privacy-preserving manner to train one model. However, when compared with conventional settings of federated learning, the proposed system differs in important ways, including (1) competition versus collaboration among the participants, and (2) assumption of the participants' intentions. 

The first difference lies in the way computational nodes work. 
In federated learning, multiple computational nodes collaborate to train a machine learning model, often by aggregating model updates after a number of iterations \cite{konen2016federated}. In comparison, the PoLe setup pits the miners against each other and let them compete for the best generationalization performance. Although collaboration arguably provides better computational efficiency, the advantage of blockchain is that it can serve as a decentralized database system, which is uniquely beneficial.  

The second major difference is the assumptions regarding the participants' intentions. Despite  some exceptions (e.g., \cite{Hitaj2017}), federated learning usually assumes the computing nodes are honest but curious \cite{Bonawitz2017} and hence do not provide explicit verification of the output of the computing nodes. In contrast, the blockchain system assumes the consensus nodes are motivated to cheat. The blockchain is designed with mechanisms to verify outputs from nodes and to maintain system integrity and consistency even when attacked.

\section{The System Architecture}

% hl {chain integrity: guaranteed by pointers from later block to the previous block}

The proposed system is a decentralized peer-to-peer network composed by two types of entities: data nodes and consensus nodes. Communication between nodes are secured via public-private asymmetric encryption schemes and happens in the form of blocks being broadcast to the entire network. 
Figure \ref{FigSystem} shows the system architecture. A data node is a user who commissions machine learning tasks via the blockchain. The consensus nodes are the suppliers of the computational power to the system; they compete to train a model that meets the requirements as specified by the data node. The winner consensus node receives the reward specified by the data node. Besides rewards from data nodes, the blockchain also functions as a decentralized data store such as those used by cryptocurrencies. In the following sections, we describe the two types of nodes in detail.

% The roles of the data provider nodes and the consensus nodes are asymmetric. Data nodes provide the monetary incentives to the consensus nodes and are assumed to be responsible users who will not abuse the mechanism. In comparison, consensus nodes can join and leave the system at any time, and are assumed to be will to cheat when possible. In the following sections, we describe the two types of nodes in detail. 
% Communication between the nodes are encrypted using public-private key systems, similar to Bitcoin. 

%The nodes with  Meanwhile, Consensus nodes can also process transactions. 
  
%We introduce our new blockchain model in this section. This model consists of two types of entities, which are data providers and consensus nodes.
%Data providers produce data and broadcast their data batch by batch to consensus nodes. hey can also send a request to the consensus node to train deep-learning models with their own uploaded data.

%The overall structure of the model is shown in Figure 3.
 %The method of generating private key and address is consistent with bitcoin.
% \begin{figure}[b]
% \includegraphics[width=0.5\textwidth]{model/blockchain-model.jpg}
% \caption{}
% \end{figure}
%Each node will be assigned a private key and address when joining the blockchain network.

 % The method of generating private key and address is consistent with bitcoin.
% \begin{figure}[b]
% \includegraphics[width=0.5\textwidth]{PAID-blockchain-network.jpg}
% \caption{}
% \end{figure}
\subsection{Data Node}
Data nodes are entities who commission machine learning tasks to the computational nodes, or consensus nodes. A task issued by a data node contains training dataset, a specification of the desired machine learning model, a minimum accuracy, and a reward. The training dataset is encrypted (see section \ref{sec:data-privacy}) and stored in blocks. A task request contains hash pointers to the training set, but not the test set. At the time of request, the task is broadcast to the entire network and added to the global task list. The global task list is an important content in the latest consensused block (previous one of the current block). The reward is immediately transferred from the data node's account to a virtual reservoir account, which will pay the winning consensus node with the best generalization performance. 

The model specification includes a full specification of network architecture such as the number and types of layers and their interconnections, but may leave out hyperparameters used for training, such as the learning rate or the weight decay coefficient. The model specification should be sufficiently detailed so that one can perform inference based on the specification from a complete set of model parameters. The specification also includes an accuracy metric, a minimum training accuracy that must be achieved, and a time limit for training. 

The test set should remain off the blockchain in order to prevent malicious consensus nodes from using the test set for training. Therefore, the data node only broadcasts the test set after it starts to receive trained models. The data node can decide to wait for a number of solutions to arrive before releasing the test set. After the test set is released, no solutions from consensus node will be accepted. This is ensured through checking against the timestamp of the test set block. 
% hl {The data nodes use digital signature to ensure that the timestamp cannot be forged.}
Data nodes will sign the timestamp to prevent forgery.
Although nodes in a distributed system may not have perfectly synchronized clocks, the time discrepancy is far smaller than the time it takes to train a machine learning model \cite{BlockDelay2019}. The test data for each task is stored in the current block for which consensus nodes have competatively contributed their computational power in training the corresponding task. With the pointers of the test dataset in the block header, any nodes receiving the new block can evaluate its validity based on the proposed block verification algorithm, and append the valid block in their blockchain based on which the node mines the next block. %a similar way to  how is the test data stored? in the current block? I don't understand that. 

%After a $enc(DB)$ is stored in a new block, a unique integer as a pointer to the location in blockchain of $enc(DB)$ will be returned by the consensus node. A data provider can retrieve all the pointers to which it uploads.

%A training model request is also called a $Task$ which is a message that contains pointers of the training data batches, objective model and rewards. The rewards are in the form of transaction containing sender address and amount of tokens. A task can be formally described as follows. %The $Mr$ can be also  and expressed as follows: 
% \begin{equation}
% Task=([Pointer_i],Model_{obj},R)
% \end{equation}
% where $[Pointer_i]$ is the list of data batch pointers; $Model_{obj}$ is the objective DL model; and $R$ is the reward message as specified 
% \begin{equation}
% R=(SAddr, Token)
% \end{equation}

% In our system the data batch $Encrypt(D_i,mpk)$ in blockchain is stored with DDH based security. The model training process is performed by block consensus nodes or miners with model parameter updated based on the inner product operations based on $Encrypt(D_i,mpk)$.  It is worthy to note that the data batches in training a model should be encrypted with the same $g$ and $r$, such that the inner product operation effects across all the data batches.
%[Revising Section 3 to this position 2020.1.19 11:30 am]

%  are The data in blockchain is theWhen data providers get the trained model, they can input the data encrypted by the same $g$ and $r$ into the model and get the predicted result in plaintext.
\subsection{Consensus Nodes and the PoLe Consensus}
\label{SectionConsensus}

Consensus nodes, also called miners, are suppliers of computational power to the network. The miners compete to perform model training tasks issued by data nodes and are rewarded as a result. 
The behavior of the consensus nodes follow the Proof-of-Learning consensus algorithm which is described as in Algorithm \ref{AlgPoLe}. 

\begin{algorithm}[!ht]
	\caption{The PoLe Consensus Algorithm}
	\label{AlgPoLe}
	\KwIn{task\_list: the task list stored in previous block\\
	\hspace{1.05cm}blk\_chain: the blockchain\\
	\hspace{1.05cm}PHS: the hash value of the previous block
		%\hspace{1.05cm} time\_max: maximum time of mining a block.\\
		%\hspace{1.05cm} required\_accuracy: accuracy required for the task;\\
	}
	%\KwOut{block\_height: the current block id\\}
	task $\leftarrow$ PopMostValuable ( task\_list ) \\
	train\_data $\leftarrow$ CollectData ( task.data\_pointers )\\
	SMLayer $\leftarrow$ CreateSMLayer(PHS)\\ %are generated according to Algorithm \ref{AlgSMLWeight};\\
	sm\_model $\leftarrow$ InsertLayer(task.model, SMLayer) \\
	received\_blks $\leftarrow$ [] \\
	\While{t $<$ time\_max}
	{
		train sm\_model for one step \\
		calculate train\_accuracy\\
		\If{train\_accuracy $\ge$ task.required\_accuracy }{
			{   blk $\leftarrow$ CreateBlock(sm\_model)\\
				broadcast blk to other consensus nodes\\
				Append(received\_blks, blk)\\
				% Return $Blk_{self}$;\\
				break \\
				%return block\_height = blk.id %$State=Blk_{self}.BlockID+1$;\\
			}
		}
		
		\If{received a new solution blk}
		{
			{ Append(blk\_chain, blk)\\
				break \\
			}	
		}
		\If{received the test data block}
		{
		    break
		}
		t++\\
	}
	blk $\leftarrow$  CreateBlock(sm\_model)\\
	broadcast blk to other consensus nodes\\
	new\_blks $\leftarrow$ WaitAndReceiveBlocks()\\
	received\_blks $\leftarrow$ Append(received\_blks, new\_blks) \\
	%produced by $SMT$ to other consensus nodes;\\
	received\_blks $\leftarrow$ Append(received\_blks, blk)\\
	sort received\_blks in descending order of test\_accuracy\\% collected blocks in descending order of accuracy;\\
	\For{each blk in received\_blks}{
		\If{VerifyBlock(blk, PHS, train\_data, test\_blk) = True}
		{ Append(blk\_chain, blk)\\
			\Return\\
		}
	}
%	$[Blk].append(Blk_{self})$;\\
%	\While{Receive new $Blk$}
%	{
%		$[Blk].append($Blk$)$;\\
%	}
%	$ValBlk = [Blk].GetValidBlock$;\\
%	Add $ValBlk$ to the end of its blockchain;\\
%	% Return $VlBlk$;\\
%	Return $State=ValBlk.BlockID+1$;\\
\end{algorithm}

When idle, a consensus node selects the task with highest value from the task list in the last block in its blockchain, which represents the best consensus known by this node, and begins training the task. The value of a task is determined by the average reward in a unit of time. Since the task list is consistent, the highest value task should be the same for all the miners. Next, it performs several maintenance steps (Line 2-4) such as collecting data according to the task description and initializing the model parameters and create the secure mapping layer (SML, see Section \ref{sec:data-privacy}) which transforms the data in ciphertext to proper input feature vectors. Since the generation of SML is related with the current block hash, the malicious nodes are unable to start mining in advance.
% hl {and prevent malicious nodes from cheating}. 

After that, the miner optimizes the specified machine learning model using a method of its choice (Line 7). When training has produced a model that meets the minimum training accuracy, the miner broadcasts a new block declaring its success (Line 8-14).  
%The data structure of a block in our system is described in in Figure \ref{blockdata}.
%We will describe the data structure of this block in detail later.}

Before a miner completes its own training, it may receive other new blocks from other miners who claim to have completed the task properly. If the miner has not received the test dataset, it saves these blocks and continues training (Line 15-18). If test data have been released before it completes the training task, the miner understands it losses the competition in this block height and terminates its training (Line 19-22). 

\begin{figure}[t]
\centering
\includegraphics[trim = 0mm 0mm 0mm 0mm,  clip, width=1\linewidth]{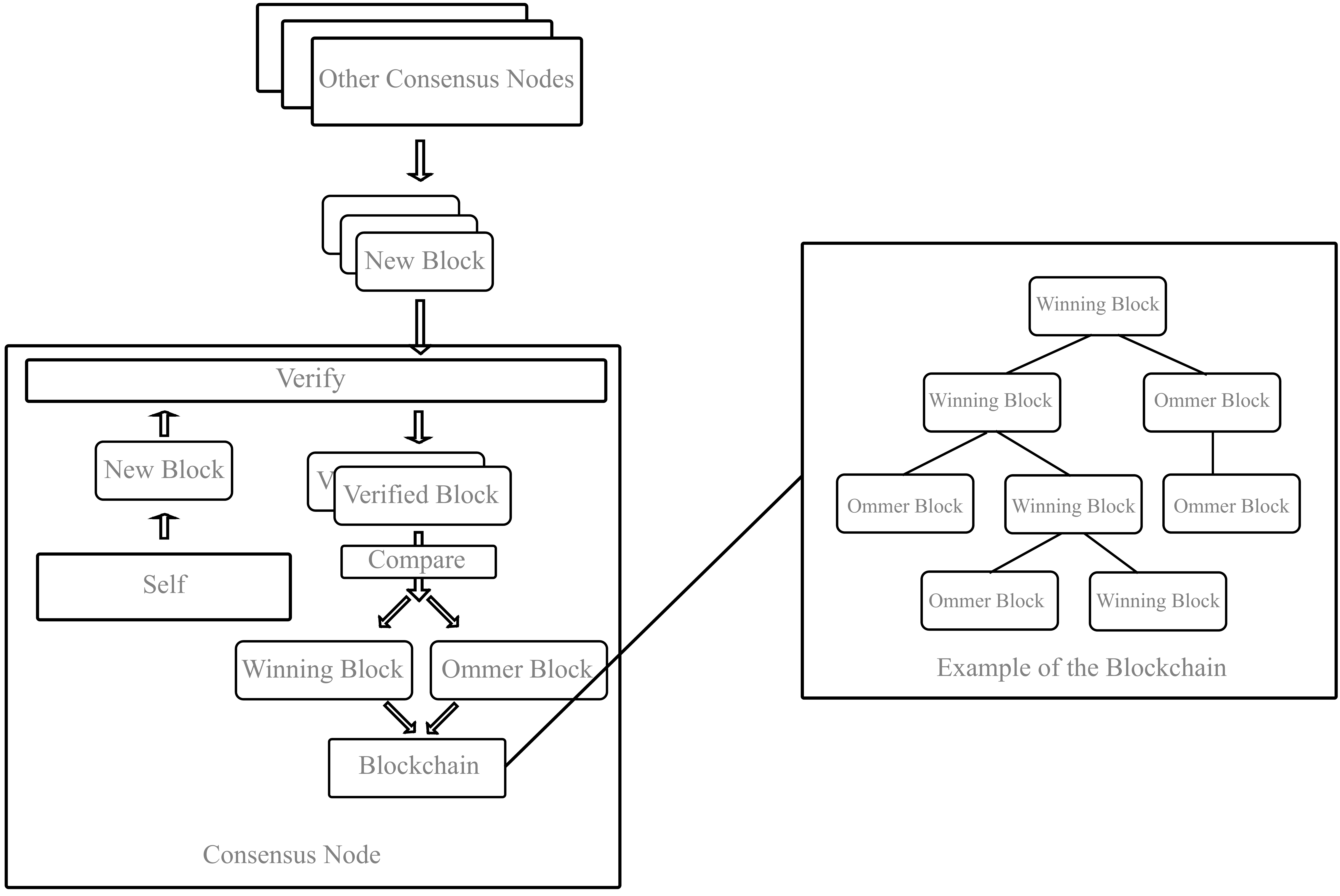}
\caption{Validating and Adding Blocks to the Blockchain}
\label{Ommerblock}
\end{figure}

The miner will terminate its training session and broadcast its new block for one of several reasons: (1) the miner has found a model that achieves the minimum training accuracy; (2) the maximum time allowed for training has been reached, or (3) miner has received the test dataset, indicating no further model solutions would be accepted.
Once a training task is completed before a test dataset is received, the trained model is packed in a new block and then broadcast to all other consensus nodes (Line 24-25).
% hl {In the training session, }
After consensus nodes receive a series of blocks and end up receiving test data (Line 26-28), they first compare the test accuracy of these blocks and sort them in a descending order of the test accuracy (Line 29). Then, miners evaluate the validity of each block according to Algorithm \ref{AlgVerification} (Line 31). 
% The verification procedure first creates a new Secure Mapping Layer according to the hash of the previous block. After that, it checks if the solution model satisifies the required training accuracy and has been completed before the test data block is released. 
A block firstly passes the verification process when Algorithm \ref{AlgVerification} returns true, then the block become the winning block(Line 31-34). The remaining blocks satisfying the requirement become ommer blocks (i.e., uncle/aunt blocks) as shown in Figure~\ref{Ommerblock}.

When a new block is accepted as a winning block, the test data is then appended in the body of the block and the rewards of the task is transferred to the block owner. The task list in the new block is formed by subtracting the completed task from the original list in the previous block and adding the newly collected tasks. Consensus nodes will consider the transactions contained in the winning block to be valid and generate new blocks by attempting the next task from the task list of the winning block. 
The whole network will only admit the transaction in the winning block.
The next winning blocks can get additional rewards by referring these ommer blocks, and the producer of the ommer block can also get rewards. The detail design of the rewards and incentives is presented in Section \ref{SectionIncentiveSecurity}

The winning block and ommer blocks are then appended to the miner's blockchain. Figure \ref{Ommerblock} {shows the whole process of collecting blocks, verifying blocks, comparing blocks and adding blocks to the blockchain.} %This round of the consensus algorithm terminates.

% For the next round, the consensus nodes will attempt the next task from the task list. 

% The training accuracy is calculated and compared with the required training accuracy of the task. Once the training accuracy is greater than the required accuracy, then a new block is created to be linked in its chain and broadcast among the network. 

% When a new block is received, the consensus node can verify the block according to Algorithm \ref{AlgVerification}. The verification procedure first creates a new Secure Mapping Layer according to the hash of the previous block. After that, it checks if the solution model satisifies the required training accuracy and has been completed before the test data block is released. The block is verified if the requirement is satisfied. 

% Algorithm \ref{AlgPoLe} takes the task list in the previous block and the current blockchain as input, and its outcome is the consistent data in distributed consensus nodes. Each consensus node chooses a task with most valuation from the body of the previous block. The valuation of a task is measured by the rewards over the expected consumed time which is estimated by the data node in its task specification.  The training dataset is retrieved and secure mapping layers (SMLayers) for training the task model is generated according to Algorithm \ref{AlgSMLWeight}. The SMLayers then serves as a decryptor to be inserted. 

Algorithm \ref{AlgVerification} shows how a received block is verified. Before a miner verifies a block, the secure mapping layer should be generated based on the previous block hash(Line 1) and insert to the trained model (Line 2). The testing accuracy is then calculated based on the released test data (Line3). If the test accuracy is greater than the required accuracy then the block pass the verification (Line 4-8). 

%if the training process terminates before time limit or hl {achieve the highest accuracy if the training continues until the maximum time}.
% When the model is trained time out, the block satisfying the minimum accuracy is treated as valid.

When the training data is completed before the specified time and the accuracy is up to the standard, if more then one block is verified at the same time, the first effective block received will be considered as the winner, and the remaining blocks will be considered as the ommer block. When the training time exceeds the maximum completion time, the block with the highest accuracy is considered as the winner, in this case, there is no ommer block.

\subsection{Data Structure of Blocks}
The block consists of header and body which is comparable with the data structure in BitCoin. %The overall structure of the block is shown in Figure} \ref{blockdata}. 
Block header contains block ID, winner's ID, selected task, previous block hash, trained model(TM), model's training accuracy , ommer block's hash and the hash of the Merkle Tree root. Block body stores data organized in a Merkle Tree, and these data include the list of uncompleted tasks in the previous block, the list of newly collected tasks, the encrypted data uploaded by the user and transactions and test data set for solved task.

%\begin{figure}[t]
%\centering
%\includegraphics[width=0.5\textwidth]{model/blockdata.png}
%\caption{The Data Structure of Blocks}
%\label{blockdata}
%\end{figure}

% \begin{algorithm}[!t]
% 	\caption{VerifyBlock (blk, PHS)}
% 	\label{AlgVerification}
% 	%\COMMENT{Verify if a new block meets accuracy or hash requirements}
% 	\KwIn{blk: received new block\\
%  	\hspace{1.1cm}test\_data: test dataset published by task requester\\
% }
% \KwOut{verified: True or False\\}
% \If{ \text{blk.train\_accuracy} $\geq$ \text{required\_accuracy}}
% 		{real\_accuracy \hspace{-1mm}$\leftarrow$\hspace{-1mm} CalAccuracy(blk.model,test\_data)\\ %Blk.TM.EvaAcc(Task.[P_{i}])$;\\
% 		\If{real\_accuracy $\geq$ required\_accuracy and blk.secure\_mapping = PHS}{
% 			\Return verified = True\\
% 		}
% 	}\Else{
% 			real\_accuracy \hspace{-1mm}$\leftarrow$\hspace{-1mm} CalAccuracy(blk.model,test\_data)\\
% 			\If{real\_accuracy $\ge$ minmal\_required\_accuracy and  blk.secure\_mapping = PHS}{
% 			\Return verified = True\\
% 		}	
% 	}
% \Return verified = False\\
% \end{algorithm}

\begin{algorithm}[!t]
	\caption{VerifyBlock (blk, PHS, train\_data)}
	\label{AlgVerification}
	\KwIn{blk: received new block\\
 	\hspace{1.1cm}train\_data: the training dataset \\
 	\hspace{1.1cm}train\_blk: the test data block \\
 	\hspace{1.1cm}PHS: the hash value of the previous block\\
}
\KwOut{verified: True or False\\}
SMLayer $\leftarrow$ CreateSMLayer(PHS)\\ %are generated according to Algorithm \ref{AlgSMLWeight};\\
	sm\_model $\leftarrow$ InsertLayer(task.model, SMLayer) \\
train\_accuracy $\leftarrow$ CalcAccuracy(sm\_model, train\_data) \\
\If{ \text{train\_accuracy} $\geq$ \text{required\_accuracy}}
		{
		\If{blk.secure\_mapping = PHS and blk.timestamp $<$ test\_blk.timestamp}{
			\Return verified = True\\
		}
	}
\Return verified = False\\
\end{algorithm}

%Consensus nodes also generate and elect new blocks using our new consensus algorithm. In the cosensus procedure, they produce deep-learning models for data providers according to received $MR$. 

%\subsection*{Transaction}
%All nodes in this network can transfer its token to another node by broadcasting a message called token transfer $TT$ to consensus nodes.
% A $TT$ includes amount of tokens, sender's address, sender's digital signature and recipient's address. It can be express as follows: 
% \begin{equation}
%TT(SenderAddr, RecipientAddr, Amount, Signature)
% \end{equation}
%
%\subsection*{Internal Data Struct of Block}

\subsection{Secure Data Storage and Secure Mapping Layers}
\label{sec:data-privacy}
We design an encryption mechansim to prevent malicious nodes from starting training before other nodes. 
We follow inner-product functional encryption~\cite{abdalla2015simple} and propose a method for the data node to upload data after encryption and the consensus nodes to access the inner product between data feature $\bm x$ and an arbitrary vector $\bm z$.

%\hl{needs more explanation of this section.}
Formally, we consider an $d$-dimensional feature vector $\bm x \in \mathbb{R}^{d}$ to be encrypted and stored. The data functional encryption and decryption process can be described in three steps:

\sloppy{\textbf{Step 1: Key generation}.  We have access to a probabilistic polynomial-time algorithm GroupGen($\lambda$), which returns a triplet $(\mathbb{G}, p, g)$. $\mathbb{G}$ is a cyclic group of order $p$ that is generated by $g \in \mathbb{G}$, where $p$ is a $\lambda$-bit prime number. }

Simply put, the cyclic group $\mathbb{G}$ is a set of $p$ elements associated with a multiplicative operation $\times$. Letting $g \times g = g^2$, the group can be written as $\mathbb{G} = \{1, g, g^2, \ldots, g^{p-1}\}$. The group is cyclic because $g^{p-1} \times g = 1$ and $1 \times g = g$. Since every element in $\mathbb{G}$ is a multiple of $g$, we say the group is generated by $g$. We let $\mathbb{Z}_p$ denote the index set $\{0, 1, \ldots, p-1\}$, which defines the range of integers that can be represented in this encryption scheme. Further, we let $\mathbb{Z}_p^D$ denote the set of $D$-dimensional vectors whose components are in $\mathbb{Z}_p$. 

The Decisional Diffie–Hellman (DDH) assumption forms the basis of this encryption scheme.
Let $\mathcal{U}(\mathbb{G})$ denote the uniform probability distribution over $\mathbb{G}$. 
The DDH assumption states that if $g^a$ and $g^b$ are i.i.d. sampled from $\mathcal{U}(\mathbb{G})$, the distribution of $g^{ab}$ is computationally indistinguishable from $\mathcal{U}(\mathbb{G})$. 

We are now ready to generate two keys, a master public key $\bm h$ used in encryption, and a master public key $\bm s$ used for decryption.
We first randomly draw $\bm s=(s_{1}, s_{2}, \ldots , s_{D})$ from $\mathbb{Z}_p^D$. The master public key is generated as $\bm h=(g^{s_{1}}, g^{s_{2}}, \ldots , g^{s_{D}}$).

\textbf{Step 2: Data encryption}. Since we can only encode integers in $\mathbb{Z}_p^D$, we first discretize the floating-point vector $\bm x \in \mathbb{R}^D$ into an integer vector $\tilde{\bm x} \in \mathbb{Z}_p^D$. 
With the master public key $\bm h$, we encrypt $ \tilde{\bm x}$ as follows. We choose a random $r \in \mathbb{Z}_p$ and compute $ct_{0}=g^r$. After that, for each component $\tilde{x}_i$ ($i \in [1,D]$) of $\bm \tilde{x}$, we compute the ciphertext $ct_{i}=h_{i}^r \cdot g^{\tilde{x}_{i}}$. At the end of this step, we have encrypted $\bm \tilde{x}$ as a $(D+1)$-dimensional ciphertext $\boldsymbol{ct}=[ct_{0}, ct_{1}, \ldots, ct_{D}]$. The ciphertext, instead of $\bm x$, is broadcast to the entire BCN. 

\textbf{Step 3: Query vector generation}. In order to maintain the data integrity of the blockchain, at every consensus node,
a set of query vectors $\{\bm z^{(i)}\}_{i=1}^I$ 
are generated from the hash of the previous block (PHS) and the Merkle root of the current block . When a 
trained NN model is being verified by other consensus nodes, the query vectors are generated again (see Algorithm \ref{AlgVerification}). Models that do not use the correct query vectors can thus be identified and discarded. 

Algorithm \ref{AlgSMLWeight} shows the detailed procedure. The query vectors are generated through switching different number of bits in the XOR operation (Line 3 of Algorithm \ref{AlgSMLWeight}). This deterministic procedure reinforces the data security of the blockchain, so that no consensus node can start training in advance. 

\textbf{Step 4: Inner-product-based decryption}. In decryption, a consensus node does not recover $\tilde{\bm x}$. Instead, the consensus node finds the inner product $\langle \tilde{\bm x}, \bm z \rangle$ for the predefined query vectors $\bm z \in \mathbb{R}^{D}$. The inner product $\langle \bm \tilde{\bm x}, \bm z \rangle$ is computed as $\log_g \left(\prod_{j \in [1,]} ct_{j}^{z_j} / ct_{0}^{\eta} \right)$, where $\bm \eta$ equals $\langle \bm s, \bm z \rangle$. 
\begin{proposition}
	The formula $\log_g \left(\prod_{j \in [1,D]} ct_{j}^{z_j} / ct_{0}^{\eta}\right)$ finds the inner product between $\bm \tilde{\bm x}$ and $\bm z$.
\end{proposition}
\emph{Proof.}
% \begin{proof}
% \renewcommand{\qedsymbol}{}
% The inner-product of weight vector $\bm z$ and plain text $\boldsymbol{x}$ is denoted by $\boldmath{p}_{i},(0\leq i\leq n)$. 
\begin{align}
\begin{split}
& \log_g\left(\prod_{j \in [1,d]} ct_{j}^{z_j} / ct_{0}^{\eta_{i}}\right) \\ 
&=\log_g\left(\prod_{j\in[1,d]}(g^{s_{j}r+\tilde{x}_{j}})^{z_{j}}/g^{r\sum_{j\in[1, d]}z_{j}s_{j}}\right) \\
&=\log_g\left(g^{\sum_{j\in[1, d]}z_{j}s_{j}r+\sum_{j\in[1,d]}z_{j}\tilde{x}_{j}-r\sum_{j\in[1,d]}z_{j}s_{j}}\right)\\ 
&=\log_g g^{\Sigma_{j\in[1, d]}z_{j} \tilde{x}_{j}}\\
&=\log_g g^{\langle \tilde{\bm x}, \boldsymbol{z} \rangle} \\
&=\langle \tilde{\bm x}, \boldsymbol{z} \rangle
\end{split}
\end{align}
% \end{proof}

\begin{figure}[t]
\centering
\includegraphics[trim = 0mm 0mm 0mm 0mm,  clip, width=1\linewidth]{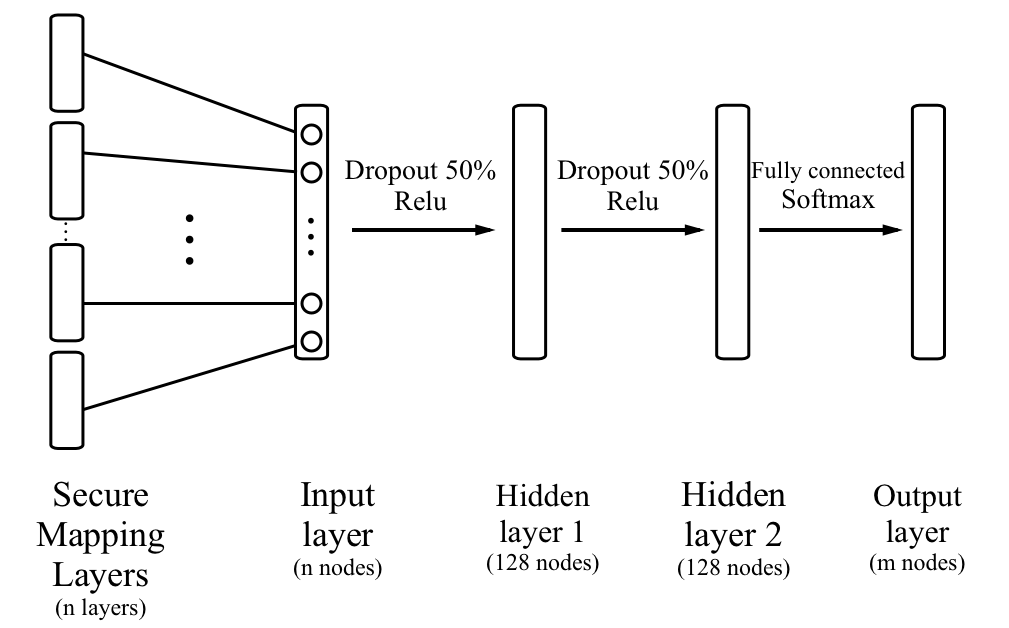}
\caption{An Example of a Neural Network Model with a Secure Mapping Layer}
\label{sml}
\end{figure}

In practice, the consensus node employs $L$ query vectors $\{\bm z^{(i)}\}_{i=1}^I$  to derive $L$ inner products, which become input features to the neural network. This is implemented as a neural network layer that is positioned before other layers and remain unchanged during training. 
Figure \ref{sml} shows an example of a neural network model with a secure mapping layer.

% We call this layer a Secure Mapping Layer. 

\textbf{Convolutional Secure Mapping Layer.} Commonly used in computer vision, convolutional neural networks (CNNs) utilize spatial information to make predictions. Visualization of network weights \cite{Zeiler2014} suggests that CNNs incrementally aggregate regional features from lower layers to create more informative features at higher layers. However, global projections in the SML layer remove spatial information in input images. This led to poor performance in preliminary experiments.  

To mitigate this problem, we propose Convolutional Secure Mapping Layer (CSML) that employs the generated query vectors as convolutional filters that are applied while sliding across the entire image. We use zero paddings around the image to maintain the height and width of the original image. Data augmentation operations like random cropping and horizontal flip can still be applied after CSML. We separately apply convolutional filters to each of the RGB channels, so that the popular color jittering augmentation can be applied using multiplicative factors on the output of CSML. More specifically, we create $L$ convolutional filters for each color channel. As a result, each color channel is transformed to $L$ channels, yielding $3L$ channels in the output.

\begin{algorithm}[t!]
	\caption{Secure Mapping Layers Generation}
	\label{AlgSMLWeight}
	\KwIn{PHS: decimal hash value\\ % of previous block\\
		\hspace{1.1cm}D: dimension of encrypted data\\
		\hspace{1.1cm}I: input dimension of the neural network\\
	\hspace{1.1cm} k: the number of bits for each weight}
	\KwOut{$\{\bm z^{(i)}\}^I_{i=1}$: the set of $I$ query vectors}
	%The vector of weights for the $ith$ secure mapping layer;\\}
	%$Cover = (len(\boldsymbol{Ct})*3) div 256$;\\
	%$Remain = len(\boldsymbol{Ct}) mod 256$;\\
	%\If{$Remain != 0$}
	%	{
	%		$Cover = Cover + 1$;\\
	%	}
	%Convert PHS to binary bits as binary(PHS);\\
	%$Size\leftarrow$ dimension of Ct;\\
	%$sum = binary(PHS)$;\\
	\For{$i$ from 1 to I}
	{ 
	    bPHS $\leftarrow$ ToBinary(PHS) \\
		primary\_weight $\leftarrow$ XOR(bPHS, bPHS $\ll i$)\\
		master\_weight $\leftarrow$ primary\_weight\\
		\While{\# bits of master\_weight $<kD$ }{
		master\_weight $\leftarrow$ Append (master\_weight,  primary\_weight)}%\\%append [$binary(PHS)$ XOR $binary(PHS)$ left shift $i$-bit]} 
	    \For{$j$ from  1 to D}{
	    $\bm{z}_{j}^{(i)}$  $\leftarrow$ ToInteger( master\_weight [j(k-1) : jk] )\\
	    }
	    %$\eta \leftarrow \langle\boldmath{s}$, %$\boldsymbol{z^{(i)}} \rangle$\\
	    %$\bm p_i \leftarrow \log_g(\prod_{j \in [1,d]} %ct_{j}^{z_{j}^{(i)}}/ct_{0}^{\eta_{i}})$
	
	}
	\Return $\{\bm z^{(1)}, \bm z^{(2)}, \ldots, \bm z^{(I)} \}$
\end{algorithm}

\subsection{Discussion: Incentives and Security}
\label{SectionIncentiveSecurity}
A blockchain based on the PoLe algorithm functions like a platform on which the data nodes post machine learning tasks for given rewards and the consensus nodes compete for them.  

The PoLe design encouranges the data node to accurately estimate the time it takes to complete the training and provide proportional reward. A data node may be tempted to intentionally overestimate the time limit in order to make consensus nodes to do more work for the same reward. However, under the current mechanism design, this will lead to low task priority. On the other hand, underestimating the training time could lead to high task priority but will cause training to terminate prematurely, yielding poor-performing models. 

Once a node generates a winning block, it will receive the reward gave by the data node and a fixed reward issued by the blockchain system $R_w$. In addition, it can also get rewards by referring previous ommer blocks. Those producers of ommer blocks can also get rewards when winning blocks refer their ommer blocks. A winning block can only reference the ommer block with the highest test accuracy. The rewards for the ommer generator is $R_r$ as calculated below. 
\begin{align}
    R_r = R_w/((H_w - H_o)*O_{num}),
    \label{EqRr}
\end{align}
Where $H_w$ stands for the height of the winning block, $H_o$ stands for the height of the ommer block and $O_{num}$ is the number of ommer blocks in the height of $H_o$.

Since all training data are available, a malicious consensus node may start training before other nodes in order to gain an unfair advantage. The SML is designed to prevent this behavior. As the SML weights are computed from the hash value of the previous block and not broadcast to the BCN, a network that does not use the correct hash cannot generate valid blocks and receive rewards. In addition, consensus nodes cannot cheat by training the model on the test data because any model that is broadcast after the release of the test data will be rejected.

% PoW algorithm let calculation of hash value contains the last calculated hash value, which makes when a node wants to tamper with imformation contained in an original block, it needs to orge all blocks after that block. The cost of this process is very high, so the blockchain network can guarantee the data security. 
In PoLe, a new block contains the hash of the previous block, which forces the attackers to recalculate all blocks after the modified block. This design increases the cost for attackers in manipulating the data in our system and enhances the system security. PoLe maps the hash of the previous block to fixed weights of secure mapping layers of secure model in current block by the Algorithm \ref{AlgSMLWeight}.
In Algorithm \ref{AlgSMLWeight}, a $k$-bit sequence is generated based on the previous block hash. This ensures that each element in $\boldsymbol{ct}$ can be assigned a $k$-bit number $\in \{0, 1, \ldots, 2^k-1\}$ from the primary weight from an XOR operation. Based on this, the weights from previous block hash can be considered to be randomly initialized. Once the previous block changes, the hash value will be different, and the trained model will be different. We assume that the most of the features of data are preserved in the results of the inner-product of data and weights, which means the performance of the two models is similar. This hypothesis is verified in the experiments in the following section. 

% In addition, for a given model and a training dataset, the amount of computation consumed to satisfy a certain training accuracy requirement is almost estimable, which enables the accuracy requirement to qualify as a proof of workload.
%\subsection{Correctness and Privacy in DL Model Training}

\section{Experiments}
The purpose of our experiments is to evaluate the efficiency and effectiveness of the proposed blockchain system with PoLe. We carry out three sets of experiments. The first experiment is designed to verify if PoLe can reliably control the amount of time between the creation of two sequential blocks in comparison with PoW. If a blockchain system can generate blocks at regular intervals, transactions carried out through the system will not experience unexpected delays, improving the smoothness of user experience.  

The second experiment compares the accuracy of deep learning model trained from a PoLe-based blockchain system and traditional method for training neural network models. This experiment aims to evaluate the effectiveness of the model trained on encrypted data and examine the training accuracy based on PoLe.%verify the assumptions made in Section \ref{SectionConsensus}. 

The third experiment is to study the effect of the same data on the accuracy when using SML inconsistent with the training. This experiment is used to study whether it is possible to forge workload by stealing other trained models and replacing them with their own SML.

\subsection{Datasets}

In these experiments we utilize three datasets, MNIST, IRIS, and CIFAR-10. The MNIST dataset represents an optical digit recognition task from black-and-white images. We follow the original split of 60,000 images for training and 10,000 images for testing. The IRIS dataset is a classification task with 3 classes and 150 data samples. We divide the data into a 90\% training set and 10\% test set. Finally, CIFAR-10 is an image classification task with 10 classes for $32\times 32$ images. The training data contain 60,000 images and the test set contains 10,000 images. 

 %that data providers send to consensus nodes. 

%system consists of two entities, one is data provider and the other is consensus node.
%Data provider produces data, encrypt data and broadcast the encrypted data to consensus nodes. They will randomly send model request to consensus nodes. Consensus nodes collects uploaded data and requests, puts them into new blocks, and elects a new block through consensus algorithm in each period.
 
\subsection{Experiment 1: Variance in Block Generation Time}

When a blockchain-based distributed ledger is used to support financial transactions, an important performance indicator is the consistency in the number of transactions that the ledger can record per unit of time. This is directly reflected in the average time it takes to generate a block and the variance in that time. A blockchain with a high variance in block generation time may lead to poor user experience because users may need to wait a long time before their trasactions are recognized by the chain. In this experiment, we create a simulation to test the variance in block generation time on PoW and PoLe chains. 
 
We build two blockchains using PoW consensus and PoLe consensus respectively. For the PoLe chain, we employ 5 data providers and 3 consensus nodes. The number of consensus nodes in PoW and our PoLe are set to be the same as 3. The expected time of generating a new block in PoW system is set to 300 seconds. Similarly, in the PoLe blockchain, we set the training accuracy threshold to 0.8 so that the exepcted time for a single model to complete training is close to 300 seconds. Both blockchains are run until 50 blocks are generated. We record the time it takes to generate every block. 

%, the minimum block production time and the average block production time of the two systems.

% \subsection*{Summary for Expriment A}
%The Figure shows the time consumption when two systems produce 20 consecutive blocks, and the overall situation is summarized in the Table. 
\begin{figure}[!h]
\centering
\includegraphics[trim = 0mm 0mm 0mm 0mm,  clip, width=1\linewidth]{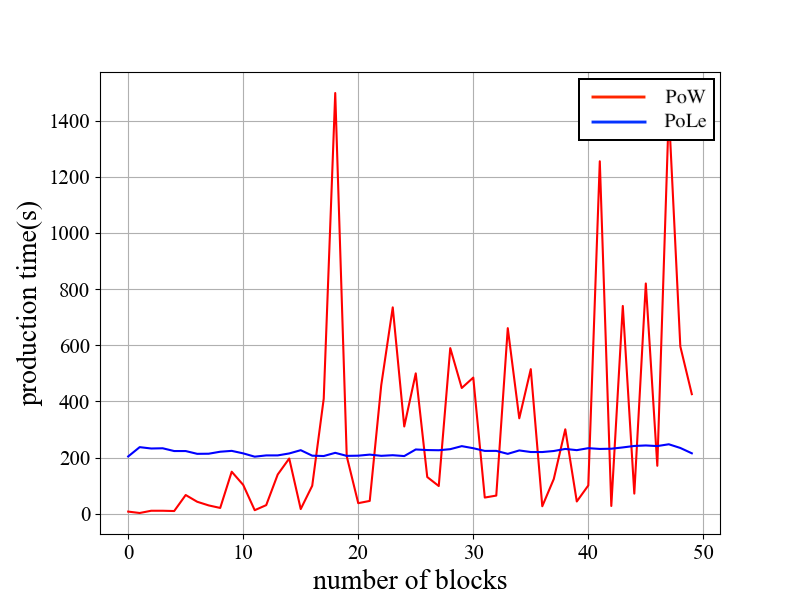}
\caption{Block Generation Time of PoW and PoLe}
\label{ExpA}
\end{figure}

 \begin{table}[h]
\centering
\caption{Statistics of Block Generation Time (in Seconds)}
\begin{tabular}{lrrrr}  
\toprule
algorithm & average   & max  & min  & variance \\
\midrule
PoLe      & 222.64  & 247.65  & 203.50  &  140.28  \\
PoW       & 293.72  & 1498.0  & 3.0     &  130548.84 \\
\bottomrule
\end{tabular}
\label{ExpAT}
\end{table}

\begin{figure*}[!ht]
\centering
\includegraphics[width=0.81\textwidth]{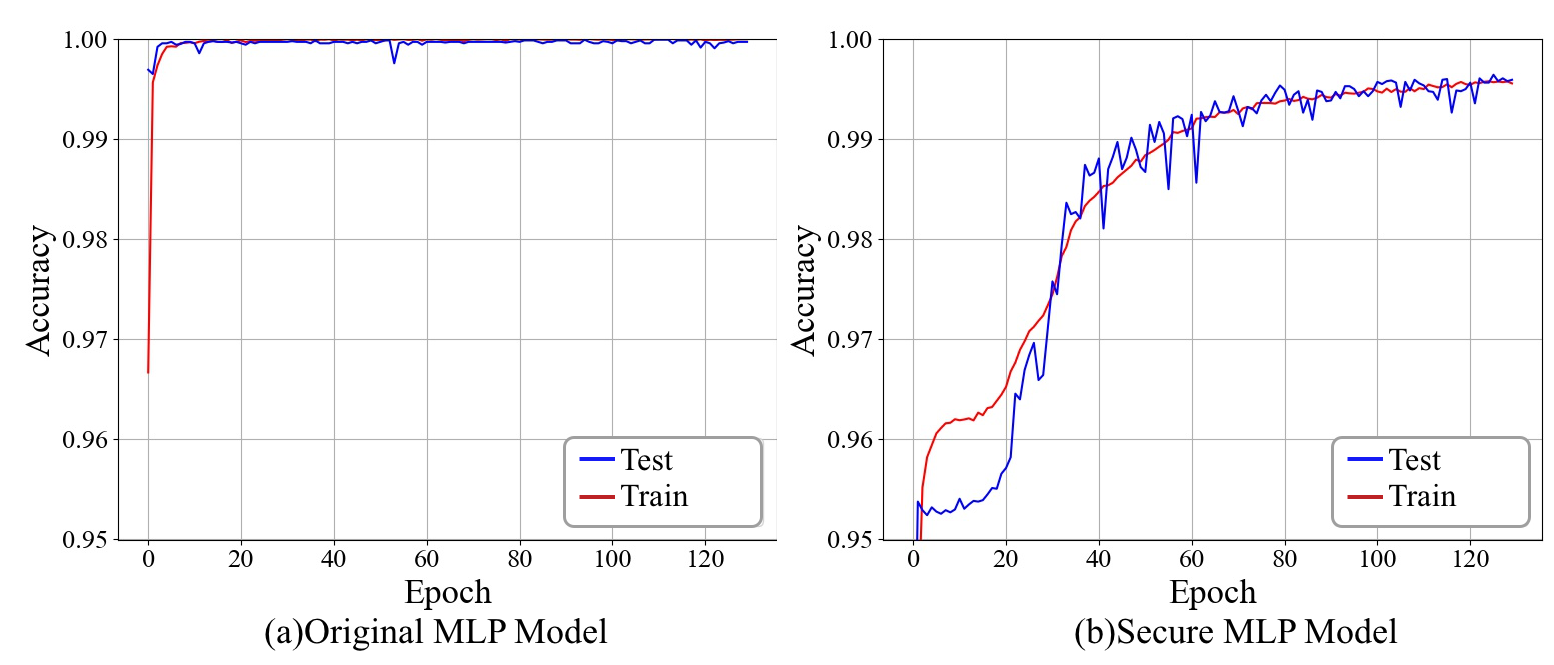}
\caption{Training and Test Accuracy Curves of the Original and the Secure MLP Models on MNIST.}
\label{Figure_B1}
\end{figure*}

\begin{figure*}[!ht]
\centering
\includegraphics[width=0.81\textwidth]{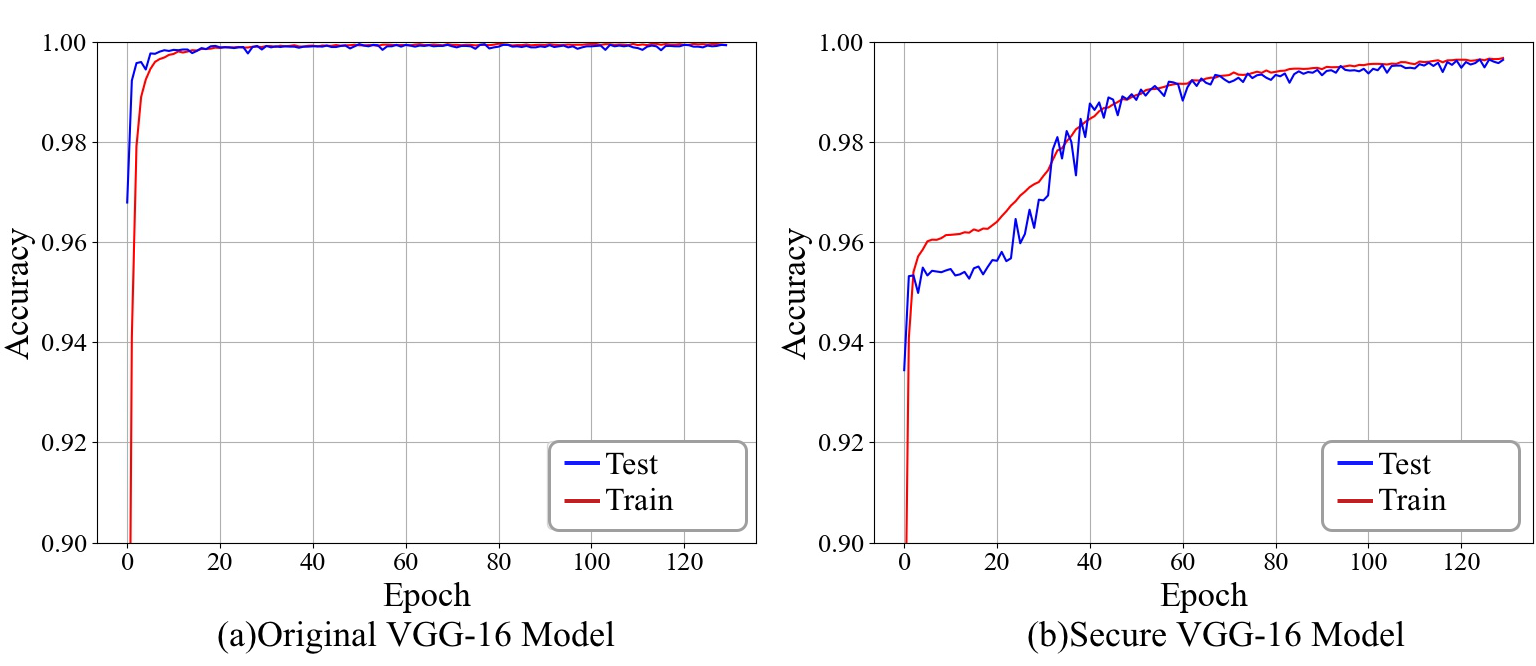}
\caption{Training and Test Accuracy Curves of the Original and the Secure VGG-16 Models on MNIST.}
\label{ExpB2}
\end{figure*}

The resulted block generation times of PoLe and PoW are shown in Figure \ref{ExpA} and the statistics are summarized in Table \ref{ExpAT}. We observe that PoLe has a much lower variance in block generation time than PoW. The reason behind this observation is that the stochastic gradient descent algorithm that optimizes the neural network is highly consistent in convergence speed, despite its apparent stochasticity. 
In contrast, due to the properities of a well-behaving hash function, whether a PoW guess of the nonce succeeds follows a high-variance distribution. 

The reader may notice that, in this simulation, we use the same machine learning problem repeatedly. We emphasize that the miners cannot reuse a network trained for previous tasks as the solution to a new task because the new task's data have been encrypted using the hash of the previous block and the Secure Mapping Layer. The feature prevents ``gun jumping'' or solution reuse and ensures the reward from the consensus is fairly allocated.

\subsection{Experiment 2: Accuracy of Secure DL Training}
We introduce the SML in order to prevent cheating behaviors. However, the SML may lower the performance of the neural network because the data have been projected onto some random directions. In this experiment, we measure the performance difference between secure networks with SMLs and unsecure networks without SMLs.

\begin{figure*}[!ht]
\centering
\includegraphics[width=0.8\textwidth]{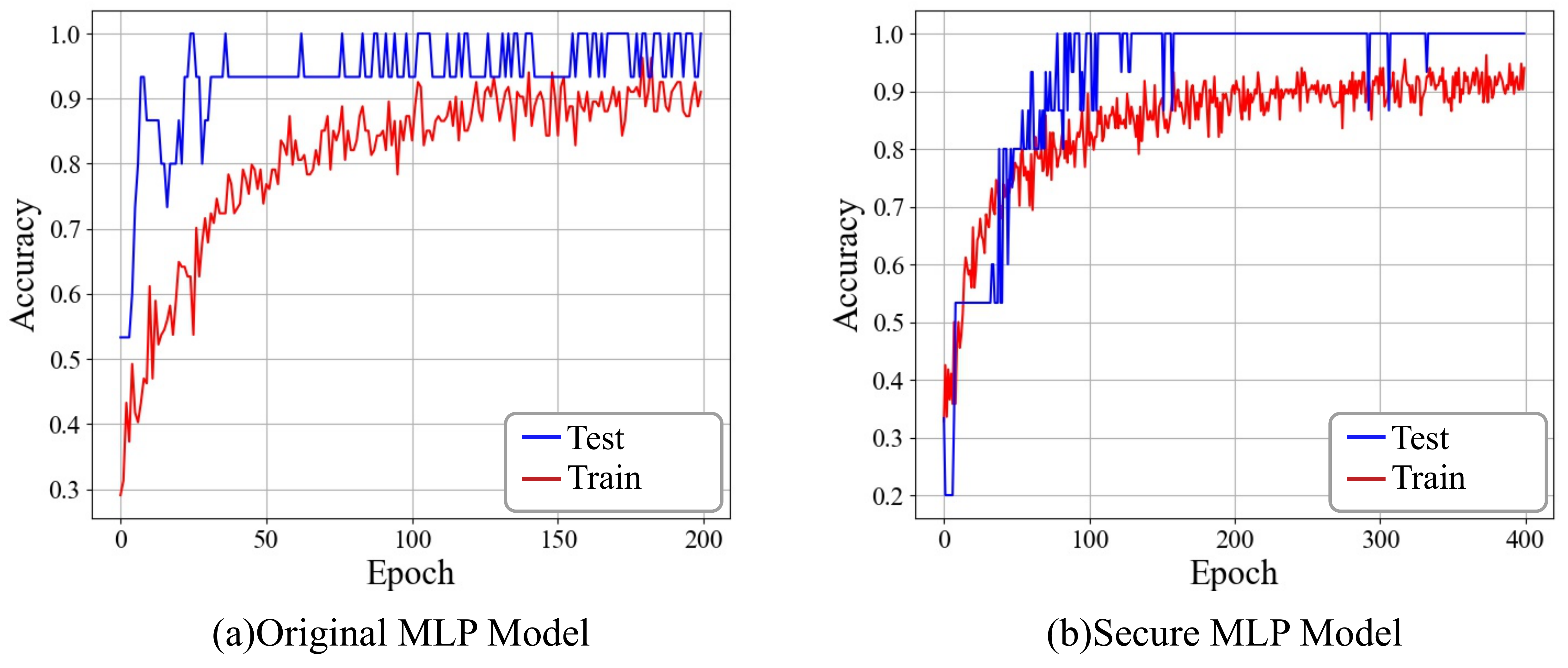}
\caption{The Accuracy Comparison of Original (Left) and Secure (Right) MLP Models for IRIS Data Set}
\label{Exp2}
\end{figure*}

\begin{figure*}[!ht]
\centering
\includegraphics[width=0.8\textwidth]{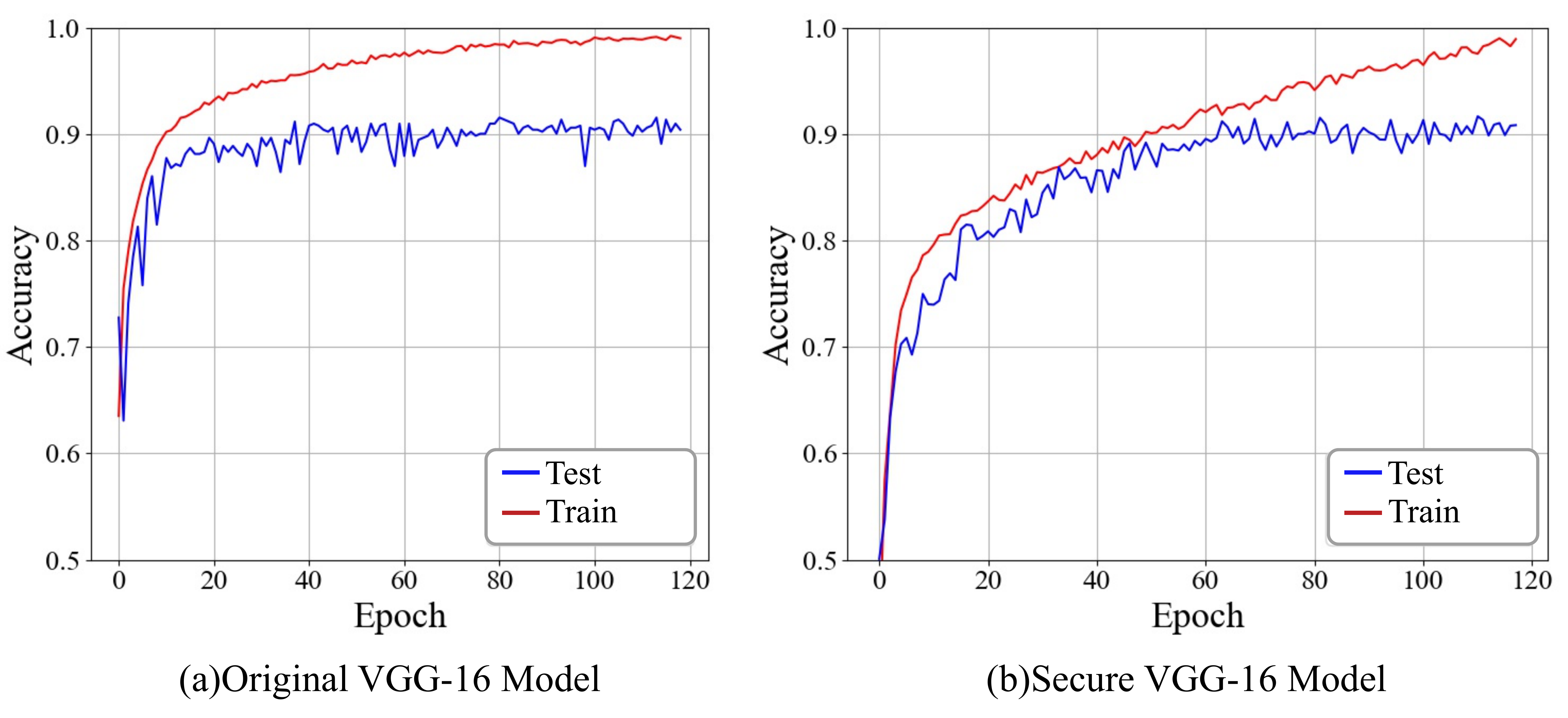}
\caption{The Accuracy Comparison of Original(Left) and Secure(Right) VGG-16 Models for CIFAR-10 Data Set}
\label{Exp3}
\end{figure*}

We create two neural networks, a multi-layer perceptron (MLP), and the VGG-16 network \cite{simonyan2014very}. For both networks, we create a secure version with a SML inserted before the first layer, and an unsecure model without SMLs. On the MNIST dataset, the value of $k$ in Algorithm \ref{AlgSMLWeight} is set to 3. The output dimension of Secure Mapping Layers is set to 128 for MLP and 256 for VGG-16. The number of query vector is set to 32 for MLP and 64 for VGG-16. In addition, we test an MLP model on the simple IRIS dataset. 
The above models are trained for 400 epochs. 
Finally, we train VGG-16 on CIFAR-10 for 120 epochs with the SML output dimension set to 2700.

% \subsection*{Summary for Experiment B}

Figures \ref{Figure_B1} and \ref{ExpB2} show the training accuracy and testing accuracy curves of MLP and VGG-16 on MNIST, respectively. Table \ref{ExpBT} reports final accuracy and the elapsed time. We observe both the secure and the unsecure models are trained to similar performance, and that the introduction of SML does not hinder model performance. We observe a small increase in the time needed for the secure models to complete the same number of epochs. The time needed to complete 400 epochs increased by 29\% for MLP and by 21\% for VGG-16. This phenomenon is expected as the SML layer brings additional computational complexity and could be mitigated with more efficient implementations. 

Figure \ref{Exp3} shows the training and test accuracy curves of the MLP on the IRIS dataset. Figure \ref{Exp2} shows the curves of VGG-16 on CIFAR-10. Similar to earlier experiments, we observe very similar performance for models with and without the SML. 

Overall, we conclude that despite the input features being projected onto different directions, the SML does not harm predictive performance significantly.

 \begin{table}[t]
\centering
\caption{Accuracy and time needed for 400 epochs of the MLP and VGG-16 models}
\begin{tabular}{lrrr}  
\toprule
model & type  & accuracy(\%) & time cost(min) \\
\midrule
MNIST MLP       & secure    & 99.90  &  14.55  \\
          & original  & 99.52  &  11.26  \\
MNIST VGG       & secure    & 99.70  &  78.56  \\
          & original  & 99.93  &  64.51  \\
CIFAR10 VGG & secure & 91.25 & 93.41 \\
        & original & 91.70 & 71.28 \\
IRIS MLP & secure & 100 & 5.14 \\
        & original & 100 & 3.25 \\
\bottomrule
\end{tabular}
\label{ExpBT}
\end{table}

\subsection{Experiment 3: Effects of Inconsistent SML on Predictive Performance}

A consensus node may be encourage to reuse a model by replacing the Secure Mapping Layer. For example, after a trained model has been broadcast to other nodes, another node may replace the SML layer and submit the solution as its own. Similarly, if a consensus node realizes the data node is re-issuing an old task, it may attempt to replace the SML in an old solution and resubmit it. Therefore, we set to empirically study the predictive accuracy of a trained model for the same data when the SML layer has been replaced. 

We use the  MNIST data set and measure the performance of the model trained in Experiment 2, as well as a the model with a replaced SML. Table ~\ref{ExpCT} shows the accuracy of the two models. We can find that the accuracy of the model for the same data set is reduced by 86.02\% when the SML is replaced. This demonstrates that replacing the SML from a model after training will result in intolerable performance degradation, which  allows the BCN to detect cheating easily.

 \begin{table}[t]
\centering
\caption{Model accuracy using original SML and replaced SML}
\begin{tabular}{lr}  
\toprule
model &  accuracy(\%)  \\
\midrule
Model + Origin SML    & 99.50\\
Model + Replaced SML      & 13.48\\
\bottomrule
\end{tabular}
\label{ExpCT}
\end{table}

\section{Conclusions}
Rapid progress in deep learning has created an unsatisfied demand for computation power. PoW-based Blockchain systems can effectively ensure data security at the cost of waiting huge computer resources. Starting from the realization that the training and testing of a machine learning model have asymmetric computational demands, we propose a new consensus algorithm called PoLe, which channels otherwise wasted compute on blockchain to the practical benefits of training machine learing models. We further designed mechanisms to prevent malicious participants from gaming the system and collecting undue rewards. Together, the proposed system achieves data integrity, defends against malicious behaviors, and does not negatively impact the predictive performance of trained models. Experimental results confirm that (1) PoLe is capable of producing a reliable stream of blocks, which support responsive data storage in a decentralized manner, and (2) the security mechanism of PoLe does not sacrifice model performance. As neural network powered AI applications and blockchain networks continue to grow in the foreseeable future, we believe the PoLe consensus mechanism will contribute to meeting their demands for compute and reducing environmental impact from energy consumption.

\bibliographystyle{ACM-Reference-Format}
\bibliography{PoDL}

%%% -*-BibTeX-*-
%%% Do NOT edit. File created by BibTeX with style
%%% ACM-Reference-Format-Journals [18-Jan-2012].

\begin{thebibliography}{27}

%%% ====================================================================
%%% NOTE TO THE USER: you can override these defaults by providing
%%% customized versions of any of these macros before the \bibliography
%%% command.  Each of them MUST provide its own final punctuation,
%%% except for \shownote{}, \showDOI{}, and \showURL{}.  The latter two
%%% do not use final punctuation, in order to avoid confusing it with
%%% the Web address.
%%%
%%% To suppress output of a particular field, define its macro to expand
%%% to an empty string, or better, \unskip, like this:
%%%
%%% \newcommand{\showDOI}[1]{\unskip}   % LaTeX syntax
%%%
%%% \def \showDOI #1{\unskip}           % plain TeX syntax
%%%
%%% ====================================================================

\ifx \showCODEN    \undefined \def \showCODEN     #1{\unskip}     \fi
\ifx \showDOI      \undefined \def \showDOI       #1{#1}\fi
\ifx \showISBNx    \undefined \def \showISBNx     #1{\unskip}     \fi
\ifx \showISBNxiii \undefined \def \showISBNxiii  #1{\unskip}     \fi
\ifx \showISSN     \undefined \def \showISSN      #1{\unskip}     \fi
\ifx \showLCCN     \undefined \def \showLCCN      #1{\unskip}     \fi
\ifx \shownote     \undefined \def \shownote      #1{#1}          \fi
\ifx \showarticletitle \undefined \def \showarticletitle #1{#1}   \fi
\ifx \showURL      \undefined \def \showURL       {\relax}        \fi
% The following commands are used for tagged output and should be
% invisible to TeX
\providecommand\bibfield[2]{#2}
\providecommand\bibinfo[2]{#2}
\providecommand\natexlab[1]{#1}
\providecommand\showeprint[2][]{arXiv:#2}

\bibitem[\protect\citeauthoryear{Abdalla, Bourse, De~Caro, and
  Pointcheval}{Abdalla et~al\mbox{.}}{2015}]%
        {abdalla2015simple}
\bibfield{author}{\bibinfo{person}{Michel Abdalla}, \bibinfo{person}{Florian
  Bourse}, \bibinfo{person}{Angelo De~Caro}, {and} \bibinfo{person}{David
  Pointcheval}.} \bibinfo{year}{2015}\natexlab{}.
\newblock \showarticletitle{Simple Functional Encryption Schemes for Inner
  Products}. In \bibinfo{booktitle}{\emph{IACR International Workshop on Public
  Key Cryptography}}. Springer, \bibinfo{pages}{733--751}.
\newblock


\bibitem[\protect\citeauthoryear{Bonawitz, Ivanov, Kreuter, Marcedone, McMahan,
  Patel, DanielRamage, Segal, and Seth}{Bonawitz et~al\mbox{.}}{2017}]%
        {Bonawitz2017}
\bibfield{author}{\bibinfo{person}{Keith Bonawitz}, \bibinfo{person}{Vladimir
  Ivanov}, \bibinfo{person}{Ben Kreuter}, \bibinfo{person}{Antonio Marcedone},
  \bibinfo{person}{H.~Brendan McMahan}, \bibinfo{person}{Sarvar Patel},
  \bibinfo{person}{DanielRamage}, \bibinfo{person}{Aaron Segal}, {and}
  \bibinfo{person}{Karn Seth}.} \bibinfo{year}{2017}\natexlab{}.
\newblock \showarticletitle{Practical Secure Aggregation for Privacy-Preserving
  Machine Learning}. In \bibinfo{booktitle}{\emph{ACM SIGSAC Conference on
  Computer and Communications Security}}.
\newblock


\bibitem[\protect\citeauthoryear{Conti, Kumar, Lal, and Ruj}{Conti
  et~al\mbox{.}}{2018}]%
        {conti2018survey}
\bibfield{author}{\bibinfo{person}{Mauro Conti}, \bibinfo{person}{E~Sandeep
  Kumar}, \bibinfo{person}{Chhagan Lal}, {and} \bibinfo{person}{Sushmita Ruj}.}
  \bibinfo{year}{2018}\natexlab{}.
\newblock \showarticletitle{A Survey on Security and Privacy Issues of
  Bitcoin}.
\newblock \bibinfo{journal}{\emph{IEEE Communications Surveys \& Tutorials}}
  \bibinfo{volume}{20}, \bibinfo{number}{4} (\bibinfo{year}{2018}),
  \bibinfo{pages}{3416--3452}.
\newblock


\bibitem[\protect\citeauthoryear{Devlin, Chang, Lee, and Toutanova}{Devlin
  et~al\mbox{.}}{2019}]%
        {Devlin2019:BERT}
\bibfield{author}{\bibinfo{person}{Jacob Devlin}, \bibinfo{person}{Ming-Wei
  Chang}, \bibinfo{person}{Kenton Lee}, {and} \bibinfo{person}{Kristina
  Toutanova}.} \bibinfo{year}{2019}\natexlab{}.
\newblock \showarticletitle{BERT: Pre-training of Deep Bidirectional
  Transformers for Language Understanding}. In
  \bibinfo{booktitle}{\emph{NAACL}}.
\newblock


\bibitem[\protect\citeauthoryear{Dinh, Liu, Zhang, Chen, Ooi, and Wang}{Dinh
  et~al\mbox{.}}{2018}]%
        {DataProcessingViewBlk2018}
\bibfield{author}{\bibinfo{person}{Tien Tuan~Anh Dinh}, \bibinfo{person}{Rui
  Liu}, \bibinfo{person}{Meihui Zhang}, \bibinfo{person}{Gang Chen},
  \bibinfo{person}{Beng~Chin Ooi}, {and} \bibinfo{person}{Ji Wang}.}
  \bibinfo{year}{2018}\natexlab{}.
\newblock \showarticletitle{Untangling Blockchain: {A} Data Processing View of
  Blockchain Systems}.
\newblock \bibinfo{journal}{\emph{{IEEE} Transactions on Knowledge and Data
  Engineering}} \bibinfo{volume}{30}, \bibinfo{number}{7}
  (\bibinfo{year}{2018}), \bibinfo{pages}{1366--1385}.
\newblock


\bibitem[\protect\citeauthoryear{Feng, Zhao, Chen, Zhao, and Zhang}{Feng
  et~al\mbox{.}}{2020}]%
        {DBLP:journals/fgcs/FengZCZZ20}
\bibfield{author}{\bibinfo{person}{Jingyu Feng}, \bibinfo{person}{Xinyu Zhao},
  \bibinfo{person}{Kexuan Chen}, \bibinfo{person}{Feng Zhao}, {and}
  \bibinfo{person}{Guanghua Zhang}.} \bibinfo{year}{2020}\natexlab{}.
\newblock \showarticletitle{Towards random-honest miners selection and
  multi-blocks creation: Proof-of-negotiation consensus mechanism in blockchain
  networks}.
\newblock \bibinfo{journal}{\emph{Future Generation Computer Systems}}
  \bibinfo{volume}{105} (\bibinfo{year}{2020}), \bibinfo{pages}{248--258}.
\newblock


\bibitem[\protect\citeauthoryear{Gervais, Karame, W{\"u}st, Glykantzis,
  Ritzdorf, and Capkun}{Gervais et~al\mbox{.}}{2016}]%
        {gervais2016security}
\bibfield{author}{\bibinfo{person}{Arthur Gervais}, \bibinfo{person}{Ghassan~O
  Karame}, \bibinfo{person}{Karl W{\"u}st}, \bibinfo{person}{Vasileios
  Glykantzis}, \bibinfo{person}{Hubert Ritzdorf}, {and} \bibinfo{person}{Srdjan
  Capkun}.} \bibinfo{year}{2016}\natexlab{}.
\newblock \showarticletitle{On the Security and Performance of Proof of work
  Blockchains}. In \bibinfo{booktitle}{\emph{Proceedings of ACM SIGSAC
  conference on computer and communications security}}. ACM,
  \bibinfo{pages}{3--16}.
\newblock


\bibitem[\protect\citeauthoryear{Hitaj, Ateniese, and Pérez-Cruz}{Hitaj
  et~al\mbox{.}}{2017}]%
        {Hitaj2017}
\bibfield{author}{\bibinfo{person}{Briland Hitaj}, \bibinfo{person}{Giuseppe
  Ateniese}, {and} \bibinfo{person}{Fernando Pérez-Cruz}.}
  \bibinfo{year}{2017}\natexlab{}.
\newblock \bibinfo{title}{Deep Models Under the GAN: Information Leakagefrom
  Collaborative Deep Learning}.
\newblock
\newblock
\showeprint[arxiv]{1702.07464}


\bibitem[\protect\citeauthoryear{Huang, Kong, Chen, Wu, Liu, and Zeng}{Huang
  et~al\mbox{.}}{2019}]%
        {huang2019towards}
\bibfield{author}{\bibinfo{person}{Junqin Huang}, \bibinfo{person}{Linghe
  Kong}, \bibinfo{person}{Guihai Chen}, \bibinfo{person}{Min-You Wu},
  \bibinfo{person}{Xue Liu}, {and} \bibinfo{person}{Peng Zeng}.}
  \bibinfo{year}{2019}\natexlab{}.
\newblock \showarticletitle{Towards Secure Industrial IoT: Blockchain System
  with Credit-based Consensus mechanism}.
\newblock \bibinfo{journal}{\emph{IEEE Transactions on Industrial Informatics}}
  (\bibinfo{year}{2019}).
\newblock


\bibitem[\protect\citeauthoryear{Kalinin and Berloff}{Kalinin and
  Berloff}{2018}]%
        {kalinin2018blockchain}
\bibfield{author}{\bibinfo{person}{Kirill~P. Kalinin} {and}
  \bibinfo{person}{Natalia~G. Berloff}.} \bibinfo{year}{2018}\natexlab{}.
\newblock \bibinfo{title}{Blockchain Platform with Proof-of-work based on
  Analog Hamiltonian Optimisers}.
\newblock
\newblock
\showeprint[arxiv]{1802.10091}


\bibitem[\protect\citeauthoryear{King and Nadal}{King and Nadal}{2012}]%
        {king2012ppcoin}
\bibfield{author}{\bibinfo{person}{Sunny King} {and} \bibinfo{person}{Scott
  Nadal}.} \bibinfo{year}{2012}\natexlab{}.
\newblock \showarticletitle{Ppcoin: Peer-to-peer Crypto-currency with
  Proof-of-stake}.
\newblock \bibinfo{journal}{\emph{self-published paper, August}}
  \bibinfo{volume}{19} (\bibinfo{year}{2012}).
\newblock


\bibitem[\protect\citeauthoryear{Konečný, McMahan, Ramage, and
  Richtárik}{Konečný et~al\mbox{.}}{2016}]%
        {konen2016federated}
\bibfield{author}{\bibinfo{person}{Jakub Konečný},
  \bibinfo{person}{H.~Brendan McMahan}, \bibinfo{person}{Daniel Ramage}, {and}
  \bibinfo{person}{Peter Richtárik}.} \bibinfo{year}{2016}\natexlab{}.
\newblock \bibinfo{title}{Federated Optimization: Distributed Machine Learning
  for On-Device Intelligence}.
\newblock
\newblock
\showeprint[arxiv]{1610.02527}~[cs.LG]


\bibitem[\protect\citeauthoryear{Larimer}{Larimer}{2014}]%
        {larimer2014delegated}
\bibfield{author}{\bibinfo{person}{D Larimer}.}
  \bibinfo{year}{2014}\natexlab{}.
\newblock \bibinfo{title}{Delegated Proof-of-stake White Paper}.
\newblock \bibinfo{howpublished}{\url{http://8btc. com/doc-view-151. html}}.
\newblock


\bibitem[\protect\citeauthoryear{Mahajan, Girshick, Ramanathan, He, Paluri, Li,
  Bharambe, and van~der Maaten}{Mahajan et~al\mbox{.}}{2018}]%
        {mahajan2018exploring}
\bibfield{author}{\bibinfo{person}{Dhruv Mahajan}, \bibinfo{person}{Ross
  Girshick}, \bibinfo{person}{Vignesh Ramanathan}, \bibinfo{person}{Kaiming
  He}, \bibinfo{person}{Manohar Paluri}, \bibinfo{person}{Yixuan Li},
  \bibinfo{person}{Ashwin Bharambe}, {and} \bibinfo{person}{Laurens van~der
  Maaten}.} \bibinfo{year}{2018}\natexlab{}.
\newblock \bibinfo{title}{Exploring the Limits of Weakly Supervised
  Pretraining}.
\newblock
\newblock
\showeprint[arxiv]{1805.00932}


\bibitem[\protect\citeauthoryear{Nakamoto}{Nakamoto}{2008}]%
        {nakamoto2008bitcoin}
\bibfield{author}{\bibinfo{person}{Satoshi Nakamoto}.}
  \bibinfo{year}{2008}\natexlab{}.
\newblock \showarticletitle{Bitcoin: A Peer-to-peer Electronic Cash System}.
\newblock \bibinfo{journal}{\emph{White Paper}} (\bibinfo{year}{2008}).
\newblock


\bibitem[\protect\citeauthoryear{Radford, Wu, Child, Luan, Amodei, and
  Sutskever}{Radford et~al\mbox{.}}{2019}]%
        {Radford2019:GPT-2}
\bibfield{author}{\bibinfo{person}{Alec Radford}, \bibinfo{person}{Jeffrey Wu},
  \bibinfo{person}{Rewon Child}, \bibinfo{person}{David Luan},
  \bibinfo{person}{Dario Amodei}, {and} \bibinfo{person}{Ilya Sutskever}.}
  \bibinfo{year}{2019}\natexlab{}.
\newblock \showarticletitle{Language Models Are Unsupervised Multitask
  Learners}. In \bibinfo{booktitle}{\emph{arXiv Preprint. arXiv:1912.12860}}.
\newblock


\bibitem[\protect\citeauthoryear{Real, Aggarwal, Huang, and Le}{Real
  et~al\mbox{.}}{2019}]%
        {Real2019:AmoebaNet}
\bibfield{author}{\bibinfo{person}{Esteban Real}, \bibinfo{person}{Alok
  Aggarwal}, \bibinfo{person}{Yanping Huang}, {and} \bibinfo{person}{Quoc~V.
  Le}.} \bibinfo{year}{2019}\natexlab{}.
\newblock \showarticletitle{Regularized Evolution Forimage Classifier
  Architecture Search}. In \bibinfo{booktitle}{\emph{AAAI}}.
\newblock


\bibitem[\protect\citeauthoryear{Ricci, Ferreira, Menasche, Ziviani, Souza, and
  Vieira}{Ricci et~al\mbox{.}}{2019}]%
        {BlockDelay2019}
\bibfield{author}{\bibinfo{person}{Saulo Ricci}, \bibinfo{person}{Eduardo
  Ferreira}, \bibinfo{person}{Daniel~Sadoc Menasche}, \bibinfo{person}{Artur
  Ziviani}, \bibinfo{person}{Jose~Eduardo Souza}, {and}
  \bibinfo{person}{Alex~Borges Vieira}.} \bibinfo{year}{2019}\natexlab{}.
\newblock \showarticletitle{Learning Blockchain Delays: A Queueing Theory
  Approach}.
\newblock \bibinfo{journal}{\emph{{ACM SIGMETRICS} Performance Evaluation
  Review}} \bibinfo{volume}{46}, \bibinfo{number}{3} (\bibinfo{year}{2019}),
  \bibinfo{pages}{122–125}.
\newblock


\bibitem[\protect\citeauthoryear{Simonyan and Zisserman}{Simonyan and
  Zisserman}{2014}]%
        {simonyan2014very}
\bibfield{author}{\bibinfo{person}{Karen Simonyan} {and}
  \bibinfo{person}{Andrew Zisserman}.} \bibinfo{year}{2014}\natexlab{}.
\newblock \showarticletitle{Very Deep Convolutional Networks for Large-scale
  Image Recognition}.
\newblock \bibinfo{journal}{\emph{arXiv preprint arXiv:1409.1556}}
  (\bibinfo{year}{2014}).
\newblock


\bibitem[\protect\citeauthoryear{So, Liang, , and Le.}{So
  et~al\mbox{.}}{2019}]%
        {So2019:EvolvedTransformer}
\bibfield{author}{\bibinfo{person}{David~R. So}, \bibinfo{person}{Chen Liang},
  \bibinfo{person}{}, {and} \bibinfo{person}{Quoc~V. Le.}}
  \bibinfo{year}{2019}\natexlab{}.
\newblock \showarticletitle{The Evolved Transformer}. In
  \bibinfo{booktitle}{\emph{Proceedings of the 36th International Conference on
  Machine Learning}}.
\newblock


\bibitem[\protect\citeauthoryear{Strubell, Ganesh, and McCallum}{Strubell
  et~al\mbox{.}}{2019}]%
        {strubell2019energy}
\bibfield{author}{\bibinfo{person}{Emma Strubell}, \bibinfo{person}{Ananya
  Ganesh}, {and} \bibinfo{person}{Andrew McCallum}.}
  \bibinfo{year}{2019}\natexlab{}.
\newblock \showarticletitle{Energy and Policy Considerations for Deep Learning
  in NLP}. In \bibinfo{booktitle}{\emph{ACL}}.
\newblock


\bibitem[\protect\citeauthoryear{Tapscott and Euchner}{Tapscott and
  Euchner}{2019}]%
        {Tapscott2019Blockchain}
\bibfield{author}{\bibinfo{person}{Don Tapscott} {and} \bibinfo{person}{Jim
  Euchner}.} \bibinfo{year}{2019}\natexlab{}.
\newblock \showarticletitle{Blockchain and the Internet of Value}.
\newblock \bibinfo{journal}{\emph{Research Technology Management}}
  \bibinfo{volume}{62}, \bibinfo{number}{1} (\bibinfo{year}{2019}),
  \bibinfo{pages}{12--18}.
\newblock


\bibitem[\protect\citeauthoryear{Yalniz, Jégou, Chen, Paluri, and
  Mahajan}{Yalniz et~al\mbox{.}}{2019}]%
        {yalniz2019billionscale}
\bibfield{author}{\bibinfo{person}{I.~Zeki Yalniz}, \bibinfo{person}{Hervé
  Jégou}, \bibinfo{person}{Kan Chen}, \bibinfo{person}{Manohar Paluri}, {and}
  \bibinfo{person}{Dhruv Mahajan}.} \bibinfo{year}{2019}\natexlab{}.
\newblock \bibinfo{title}{Billion-scale Semi-supervised Learning for Image
  Classification}.
\newblock
\newblock
\showeprint[arxiv]{1905.00546}


\bibitem[\protect\citeauthoryear{Yang, Liu, Chen, and Tong}{Yang
  et~al\mbox{.}}{2019}]%
        {yang2019federated}
\bibfield{author}{\bibinfo{person}{Qiang Yang}, \bibinfo{person}{Yang Liu},
  \bibinfo{person}{Tianjian Chen}, {and} \bibinfo{person}{Yongxin Tong}.}
  \bibinfo{year}{2019}\natexlab{}.
\newblock \showarticletitle{Federated Machine Learning: Concept and
  Applications}.
\newblock \bibinfo{journal}{\emph{ACM Transactions on Intelligent Systems and
  Technology}} \bibinfo{volume}{10}, \bibinfo{number}{2}
  (\bibinfo{year}{2019}).
\newblock


\bibitem[\protect\citeauthoryear{Zeiler and Fergus}{Zeiler and Fergus}{2014}]%
        {Zeiler2014}
\bibfield{author}{\bibinfo{person}{Matthew~D. Zeiler} {and}
  \bibinfo{person}{Rob Fergus}.} \bibinfo{year}{2014}\natexlab{}.
\newblock \showarticletitle{Visualizing and Understanding Convolutional
  Networks}. In \bibinfo{booktitle}{\emph{ECCV}}. \bibinfo{publisher}{Springer
  International Publishing}, \bibinfo{pages}{818--833}.
\newblock


\bibitem[\protect\citeauthoryear{Zhou, Dou, , and Li}{Zhou
  et~al\mbox{.}}{2019}]%
        {Xin2019}
\bibfield{author}{\bibinfo{person}{Xin Zhou}, \bibinfo{person}{Dejing Dou},
  \bibinfo{person}{}, {and} \bibinfo{person}{Boyang Li}.}
  \bibinfo{year}{2019}\natexlab{}.
\newblock \showarticletitle{Searching for Stage-wise Neural Graphs in the
  Limit}. In \bibinfo{booktitle}{\emph{arXiv Preprint. arXiv:1912.12860}}.
\newblock


\bibitem[\protect\citeauthoryear{Zhuang, Liu, Chen, and Ai}{Zhuang
  et~al\mbox{.}}{2019}]%
        {PoRyuan2019}
\bibfield{author}{\bibinfo{person}{Qianwei Zhuang}, \bibinfo{person}{Yuan Liu},
  \bibinfo{person}{Lisi Chen}, {and} \bibinfo{person}{Zhengpeng Ai}.}
  \bibinfo{year}{2019}\natexlab{}.
\newblock \showarticletitle{Proof of Reputation: A Reputation-based Consensus
  Protocol for Blockchain Based Systems}. In
  \bibinfo{booktitle}{\emph{Proceedings of the 2019 International Electronics
  Communication ConferenceJuly}}. \bibinfo{pages}{131--–138}.
\newblock


\end{thebibliography}

\end{document}